# Thickness dependence of antiferromagnetic phase transition in Heisenberg-type MnPS$_3$

Soo Yeon Lim[a], Kangwon Kim[a], Sungmin Lee[b,c], Je-Geun Park[b,c], and Hyeonsik Cheong[a]

[a]Department of Physics, Sogang University, Seoul 04107, Korea

[b]Department of Physics and Astronomy, Seoul National University, Seoul 08826, Korea

[c]Center for Quantum Materials, Seoul National University, Seoul 08826, Korea

[*]E-mail: hcheong@sogang.ac.kr

**ABSTRACT**

The behavior of 2-dimensional (2D) van der Waals (vdW) layered magnetic materials in the 2D limit of the few-layer thickness is an important fundamental issue for the understanding of the magnetic ordering in lower dimensions. The antiferromagnetic transition temperature $T_N$ of the Heisenberg-type 2D magnetic vdW material $MnPS_3$ was estimated as a function of the number of layers. The antiferromagnetic transition was identified by temperature-dependent Raman spectroscopy, from the broadening of a phonon peak at 155 cm$^{-1}$, accompanied by an abrupt redshift and an increase of its spectral weight. $T_N$ is found to decrease only slightly from ~78 K for bulk to ~ 66 K for 3L. The small reduction of $T_N$ in thin $MnPS_3$ approaching the 2D limit implies that the interlayer vdW interaction is playing an important role in stabilizing magnetic ordering in layered magnetic materials.

1. Introduction

2-dimensional (2D) van der Waals (vdW) magnetic materials have recently attracted much interest owing to interesting physical properties depending on magnetic phases and structures [1] as well as potential applications in flexible or nanoscale spintronic devices. From the studies of 2D ferromagnetic materials such as $CrI_3$ [2] and $CrBr_3$ [3], interesting effects of the interlayer interaction on the magnetic transition have been reported. For example, bulk $CrI_3$ is known to be an Ising-type ferromagnet, but few-layer $CrI_3$ exhibits interesting layer-number-dependence: ferromagnetic for odd number of layers and antiferromagnetic for even number of layers [2]. On the other hand, bilayer (2L) $CrBr_3$ exhibits ferromagnetic ordering for 2H stacking and antiferromagnetic ordering for 3R stacking whereas monolayer (1L) $CrBr_3$ is ferromagnetic[3]. Furthermore, the Curie temperature ($T_C$) of few-layer $Cr_2Ge_2Te_6$ is suppressed as the number of layers decreases although exact determination of $T_C$ was difficult in the initial magneto-optic measurements [4] . For 2D vdW antiferromagnetic materials, direct measurements of the magnetic ordering are not easy because of the lack of net magnetization and the small sample volume. Raman spectroscopy can be used when a certain phonon mode couples with the magnetic ordering to show abrupt changes in the line shape, the frequency, or the intensity [5–12]. Sometimes, the magnon signals that reflect the magnetic structure are observed in the Raman spectrum and can be correlated with the magnetic ordering [7,13].

Transition-metal phosphorus trisulfides (TMPS$_3$; TM= Fe, Ni, and Mn) are one of 2D vdW antiferromagnet families with the bandgap energies in the range of 1.5 to 3.0 eV [14]. Although the crystal structure is identical, the type of magnetic ordering depends on the transition metal element: Ising type for FePS$_3$, XXZ type for NiPS$_3$, and Heisenberg type for MnPS$_3$ [14–16]. Ising-type FePS$_3$ exhibits antiferromagnetic ordering down to the 1L limit with almost no change in the Néel temperature ($T_N$) [5,6]. For XXZ-type NiPS$_3$, which behaves like an XY-system [17,18] at low temperatures, exhibit antiferromagnetic ordering down to 2L with a slight decrease of $T_N$ as the thickness is decreased, but the magnetic ordering is suppressed in 1L [7]. These behaviors are consistent with the XY (XXZ) system being more prone to fluctuations than the Ising system in the 2D limit. Heisenberg-type MnPS$_3$ also shows antiferromagnetic ordering at low temperatures, and an indication of the ordering was observed down to 2L. However, the dependence of $T_N$ on the number of layers has not been studied systematically due to the very weak signal of the Raman mode that reflects the magnetic ordering. Because a Heisenberg system is more susceptible to fluctuations, in a simple picture, it is expected that $T_N$ will decrease rapidly as the number of layers decreases. However, we found that $T_N$ decreases as the number of layers decreases, but the decrease is modest, not more than the case of XXZ-type NiPS$_3$.

## 2. Methods

Single crystals of $MnPS_3$ were grown by a self-flux chemical vapor transport method in a quartz tube ampoule evacuated to high vacuum. The details of the crystal growth have been published previously [9]. Few-layer $MnPS_3$ samples were mechanically exfoliated from the single crystal onto $SiO_2$/Si substrates with 90 nm oxide layer. The samples on substrates were kept in vacuum to avoid degradation from air exposure.

We carried out Raman scattering measurements with an Ar-ion laser with the wavelength of 488 nm (2.54 eV) and the power of ~50 µW. A 40× objective lens (N.A.=0.6) was used to focus the laser on the samples with a spot of ~1 µm in diameter and to collect Raman scattered photons from the samples. A substrate with exfoliated samples was loaded into a He-flow optical cryostat (Oxford MicrostatHe2). The scattered light from the sample was dispersed with a Jobin-Yvon Horiba iHR550 spectrometer (2400 grooves/mm) and was detected with a CCD using liquid nitrogen for cooling. A spike filter was used to removed unwanted plasma lines from the laser, and a long-pass edge filter was used to block the Rayleigh scattered light from entering the spectrometer. The cutoff of the edge filter was ~100 $cm^{-1}$. Polarized Raman measurements were performed with polarizers and λ/2 wave plates to obtain Raman signals selectively in appropriate polarization configurations of incident and scattered light. The sample was kept in vacuum during Raman measurements to avoid photochemical degradation due to laser exposure.

## 3. Results

### 3.1. Crystal structures and sample images of layered $MnPS_3$

Figure 1(a) shows the crystal structure of a monolayer of $MnPS_3$ having a hexagonal structure (point group of $D_{3d}$). Bulk $MnPS_3$ has the monoclinic structure (point group of $C_{2h}$) with successive layers stacked with an oblique angle. The Mn ions hosting spins constitute a honeycomb lattice whereas P and S atoms surround the Mn-hexagons like a dumbbell. Figure 1(b) shows the spin alignment of Mn ions when the temperature is below $T_N$. Each Mn ion is coupled to its three nearest neighbors antiferromagnetically, leading to antiferromagnetic ordering within the monolayer unit.

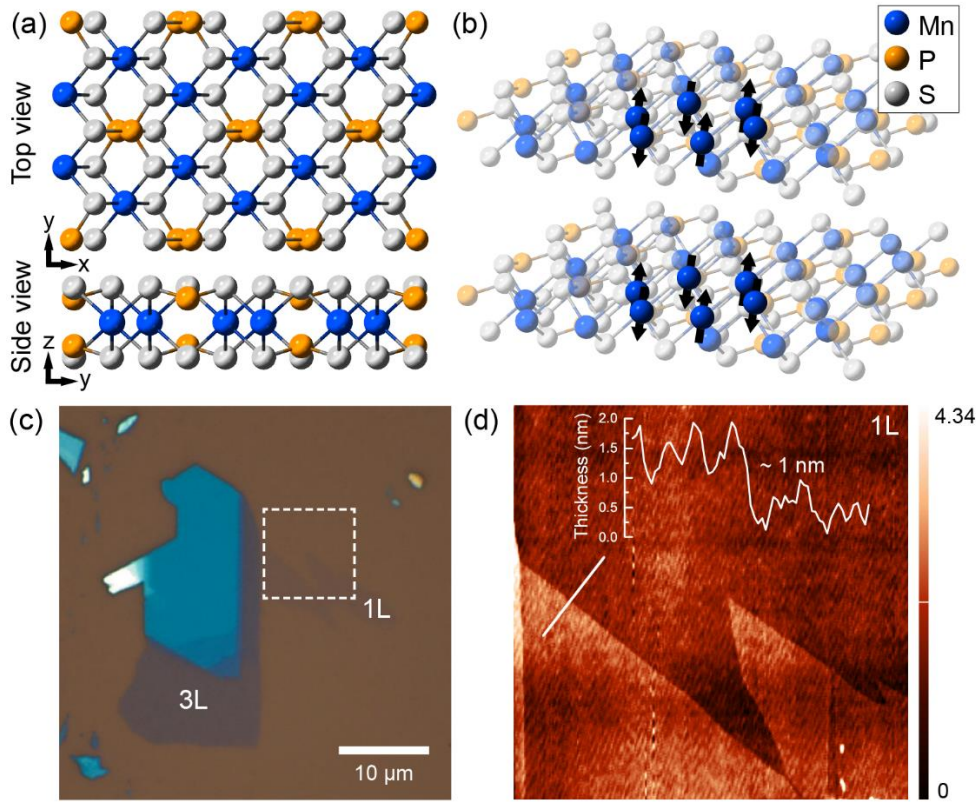

**Fig. 1.** (a) Crystal structures of 1L MnPS$_3$ in top and side views. (b) Spin alignment of Mn ions below $T_N$. The black arrows indicate the spin direction of the Mn atoms. (c) Optical image of an exfoliated MnPS$_3$ sample on a SiO$_2$/Si substrate. The dashed-white box indicates the region where AFM measurements were performed. (d) AFM topography image of 1L MnPS$_3$ with ~ 1 nm thickness. The cross section profile was taken along the white line.

Figures 1(c) and (d) show an optical microscope image and an atomic force microscope image (AFM), respectively, of a typical sample. Although 1L MnPS$_3$ is barely seen in an optical image as shown in Fig. 1(c), one can clearly recognize the 1L MnPS$_3$ flake in the AFM topography image of Fig. 1(d). The thickness of 1L MnPS$_3$ was measured from AFM to be ~ 1 nm, which is consistent with the layer spacing of ~0.7 nm measured by X-ray diffraction [19].

*3.2. Raman spectra of layered MnPS$_3$ at room temperature*

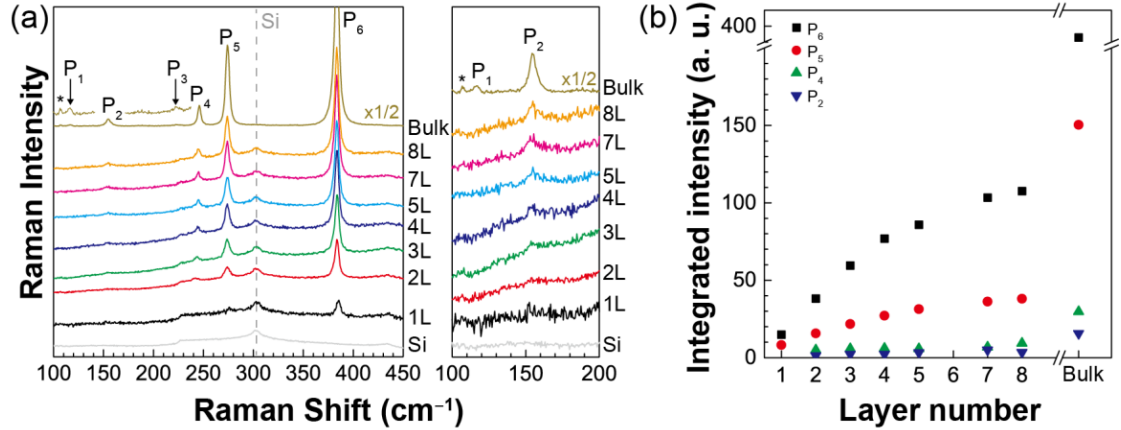

**Fig. 2.** (a) Raman spectra of 1L to 8L and bulk MnPS$_3$ samples. The gray dashed line indicates a Raman peak from the substrate. The asterisk mark indicates the laser plasma line that was not fully filtered. (b) Integrated intensity for P$_2$, P$_4$, P$_5$, and P$_6$ peaks as a function of the layer number.

Figure 2(a) shows the Raman spectra of MnPS$_3$ samples from 1L to 8L and bulk, measured with the 488-nm laser line as the excitation source in the parallel polarization configuration [$\bar{z}(xx)z$]. Whereas six Raman modes, P$_1$ through P$_6$, are observed from the bulk sample, the weak peaks P$_1$ at ~120 cm$^{-1}$ and P$_3$ at ~230 cm$^{-1}$ were not resolved in the spectra from few-layer MnPS$_3$. The intensities of all the Raman modes gradually decrease as the layer number decreases as shown in Fig. 2 (b). Although the sample was kept in vacuum at all times, we noticed that some degradation occurred during Raman measurements even with the low excitation laser power of ~50 µW if the laser illuminates the same spot for an extended period of 500 s or longer (See Fig. S1 for details). In order to avoid such degradation, we obtained Raman spectra from multiple points within an area of ~10×8

μm$^2$ with a short integration time (50 or 100 s) at each point and used the average of such spectra for better signal-to-noise ratios. The AFM topography image and Raman mapping show that the sample is highly uniform over the measured area. (See Fig. S2 and Fig. S3).

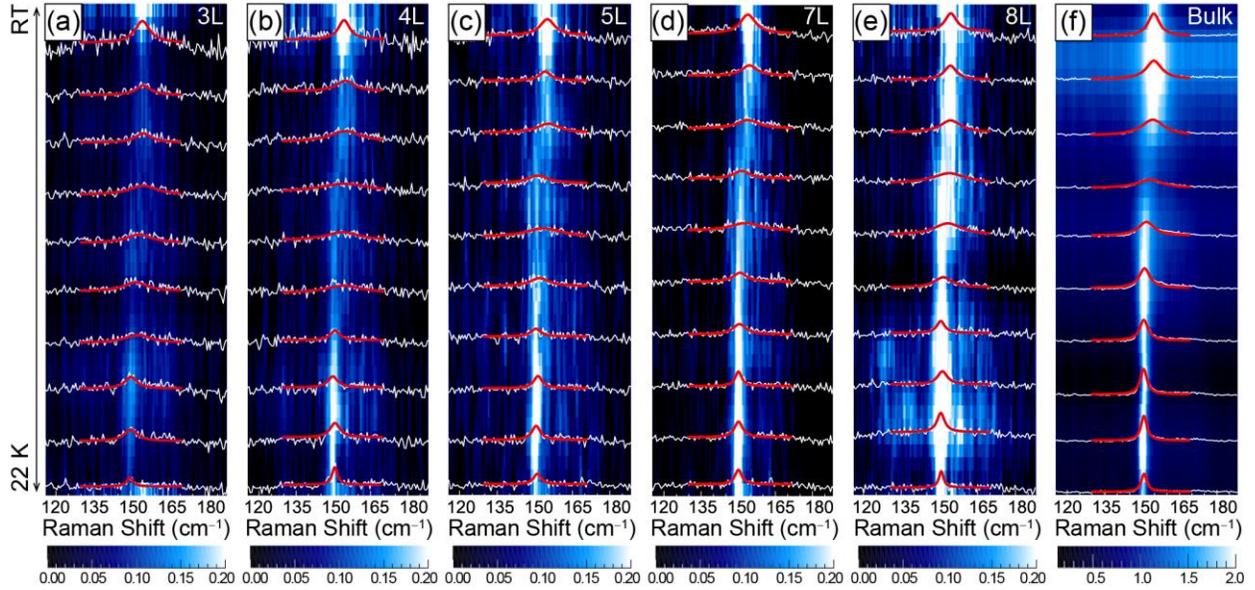

**Fig. 3.** Raman spectra measured at several representative temperatures (22, 32, 38, 44, 50, 56, 62, 72, 82, and 290 K) with Raman intensity plots at all measured temperatures from 22 K to 290 K of (a) 3L, (b) 4L, (c) 5L, (d) 7L, (e) 8L, and (f) bulk MnPS$_3$. The red curves are fitting curves for P$_2$.

*3.3. Temperature dependence of P$_2$ in few-layer MnPS$_3$*

The antiferromagnetic transition temperature $T_N$ was estimated from the temperature dependence of the Raman spectra. In previous work, it was found that the peak P$_2$, which comes from vibrations of Mn ions [20], can be used as the indicator of antiferromagnetic ordering; the peak redshifts abruptly as the temperature is lowered through $T_N$ and the linewidth increases near $T_N$ [9]. The red shift is attributed to the coupling of the phonon vibration modes with the ordered magnetic moments below the $T_N$. Furthermore, large fluctuations near the phase transition temperature should disrupt the coherence of the lattice vibration modes, shortening the phonon lifetime, which in turn causes broadening of the Raman peak near $T_N$ [9]. Figure 3 shows the Raman spectra of 3-, 4-, 5-, 7-, 8L and bulk MnPS$_3$ measured at several representative temperatures. Unfortunately, the P$_2$ signal from 1L and 2L samples were too weak to be resolved at low temperatures. We used the cross polarization configuration [$\bar{z}(xy)z$] for this set of measurements to better resolve P$_2$; the weak luminescent background can be suppressed with the cross polarization whereas P$_2$ has the same intensity in parallel and cross polarizations. As the temperature is lowered toward the magnetic transition temperature, the P$_2$ peak broadens and becomes weak. As the temperature is lowered even more, this peak re-emerges at a lower frequency with a sharper line shape. This trend is common to all the thicknesses measured. For example, the linewidth of P$_2$ for exfoliated bulk MnPS$_3$ has a maximum near ~ 78 K, which is the $T_N$ of single crystal MnPS$_3$ [9,10].

### 3.4. Néel temperature of few-layer MnPS$_3$

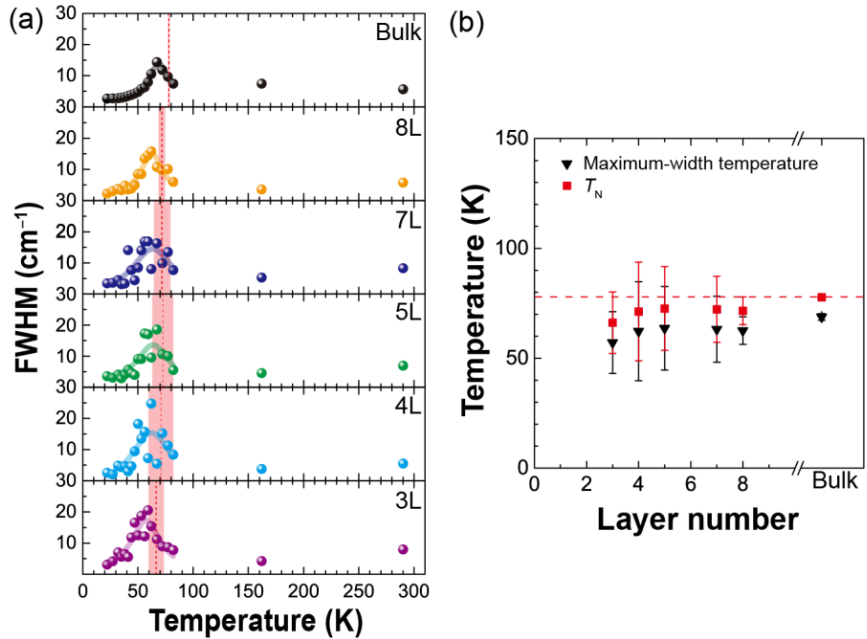

**Fig. 4.** (a) Temperature dependence of FWHM of P$_2$ from Raman spectra in Fig. 3. The red-dashed lines indicate the estimated $T_N$, located at 9 K above the peak of FWHM, and the shaded bands indicate the experimental uncertainties.
8

(b) Maximum-width temperature and $T_N$ as a function of the number of layers. The error bars correspond to the linewidth of the Lorentzian fit curves in (a).

In order to estimate $T_N$ for each sample, we analyzed the line shape by fitting each spectrum with a Lorentzian function. Figure 4(a) shows the FWHM of $P_2$ for each sample extracted from fitting. Since the local temperature at the sample can be different from the reading from the thermometer of the cryostat, we re-calibrated the temperature scale by comparing the temperature dependence of the Raman spectrum of the bulk sample, with our previous measurements on a bulk single crystal measured in a He closed cycle cryostat. The FWHM of all the samples increase sharply in the temperature range 50 to 70 K. In order to estimate the transition temperature $T_N$ for each sample, we fit the experimental data for FWHM to a single Lorentzian function. In the case of bulk $MnPS_3$, the maximum of the FWHM is located 9 K below $T_N$. Although this offset between the maximum of FWHM and $T_N$ may depend on the number of layers, we assume a constant offset as an approximation. The estimated $T_N$ for each sample is indicated by a vertical broken line with the uncertainty given by the width of the shaded vertical band. Figure 4(b) summarizes the dependence of $T_N$ on the number of layers. It shows that $T_N$ decreases only slightly as the number of layers decreases: the $T_N$ of 3L is about 12 K lower than that of bulk.

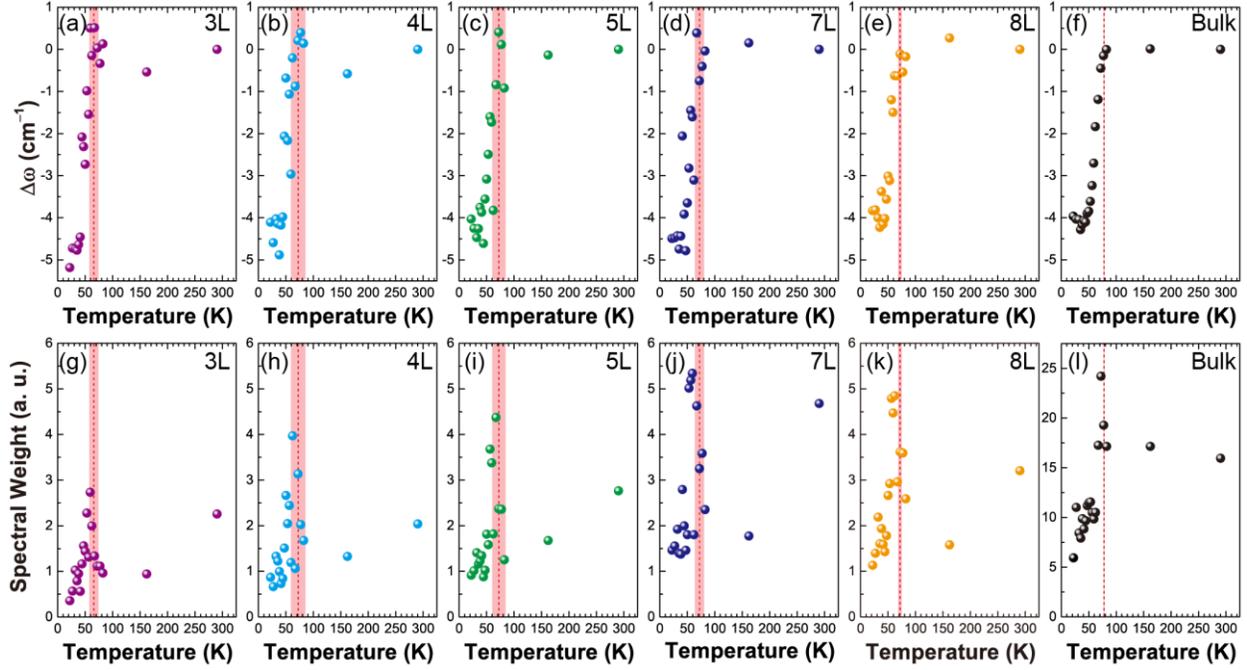

**Fig. 5.** Temperature dependence of (a-f) peak shift and (g-l) spectral weight of the $P_2$. The peak shifts are adjusted for the anharmonic shift of the phonons due to thermal effect. The red-dashed lines and the shaded bands indicate $T_N$ and its uncertainty from Fig. 4.

In order to crosscheck the estimated $T_N$, we analyzed other spectral parameters. Figure 5 shows the relative peak shift ($\Delta\omega$) and the spectral weight of $P_2$ for each sample. The peak position is adjusted for the overall shift of phonon frequencies due to the anharmonicity [21]. The transition temperature $T_N$ and its uncertainty from Fig. 4 are also shown. One can clearly see that the peak abruptly redshifts as the temperature is lowered through $T_N$ whereas other peaks blueshift (see Fig. S4), and the spectral weight is increased near $T_N$. These results are compatible with the estimate of $T_N$ in Fig. 4.

Our results are consistent with another work in which $T_N$ was estimated to be independent of the number of layers down to 5L [8]: the shift in $T_N$ between bulk and 5L is less than 5 K in our data. The ~12 K shift of $T_N$ between bulk and 3L is of the same order as in the case of XXZ-type $NiPS_3$ [7], which is somewhat surprising. In a Heisenberg-type system, the magnetic ordering should not exist in the 1L limit according to the Mermin-Wagner theorem [22], and $T_N$ is expected to approach 0 K rapidly as the thickness is reduced. Our results indicate that 3L $MnPS_3$ is closer to bulk than to the 2D limit of 1L. This implies that the interlayer interaction is playing an important role in stabilizing the magnetic ordering although the overall dimensionality is close to the 2D limit. In this respect we should note on a recent report in which $T_N$ of $MnPS_3$ was estimated from the temperature dependence of the magnetoresistance [23]. Although this estimate was rather indirect, the authors claimed that $T_N$ does not change down to 1L and that the Mermin-Wagner theorem does not apply to $MnPS_3$ due to long-range interactions. More follow-up studies are needed to elucidate the intricate interplay of dimensionality and interlayer interactions in Heisenberg-type layered vdW magnetic materials.

## 4. Conclusions

The thickness dependence of the antiferromagnetic transition temperature $T_N$ of the Heisenberg-type 2D magnetic vdW material $MnPS_3$ was estimated from temperature dependence of the Raman spectra. The antiferromagnetic transition was identified by the broadening of a phonon peak $P_2$, accompanied by an abrupt redshift and an increase of the spectral weight. We found that $T_N$ decreases only slightly from ~78 K for bulk to ~ 66 K for 3L. The unexpected small reduction of $T_N$ in thin $MnPS_3$ implies that the interlayer vdW interaction is playing an important role in stabilizing magnetic ordering in layered magnetic materials.

**Declaration of competing interest**

The authors declare no competing financial interest.

**Acknowledgments**

This work was supported by the National Research Foundation (NRF) grant funded by the Korean government (MSIT)(2019R1A2C3006189; 2017R1A5A1014862, SRC program: vdWMRC center; 2020R1A3B2079375, Leading Researcher Program). S.Y.L. acknowledges support from Hyundai Motor Chung Mong-Koo Foundation.

**Appendix A. Supplementary data**

Supplementary data associated with this article can be found, in the online version, at http://dx.doi.org/xx.xxx.

# Supporting Information

# Thickness dependence of antiferromagnetic phase transition in Heisenberg-type MnPS$_3$

Soo Yeon Lim[a], Kangwon Kim[a], Sungmin Lee[b,c], Je-Geun Park[b,c], and Hyeonsik Cheong[a,*]

[a] Department of Physics, Sogang University, Seoul 04107, Korea

[b] Department of Physics and Astronomy, Seoul National University, Seoul 08826, Korea

[c] Center for Quantum Materials, Seoul National University, Seoul 08826, Korea

*E-mail: hcheong@sogang.ac.kr-

**Contents:**

- **Fig. S1.** Laser-induced degradation of few-layer MnPS$_3$.
- **Fig. S2**. Optical and AFM images of 3L MnPS$_3$
- **Fig. S3**. Raman spectra and maps in terms of peak position, integrated intensity, and FWHM of P$_6$ in 3L MnPS$_3$ at 27K.
- **Fig. S4**. Temperature-dependent peak positions of P$_4$, P$_5$, and P$_6$ of 3L MnPS$_3$.

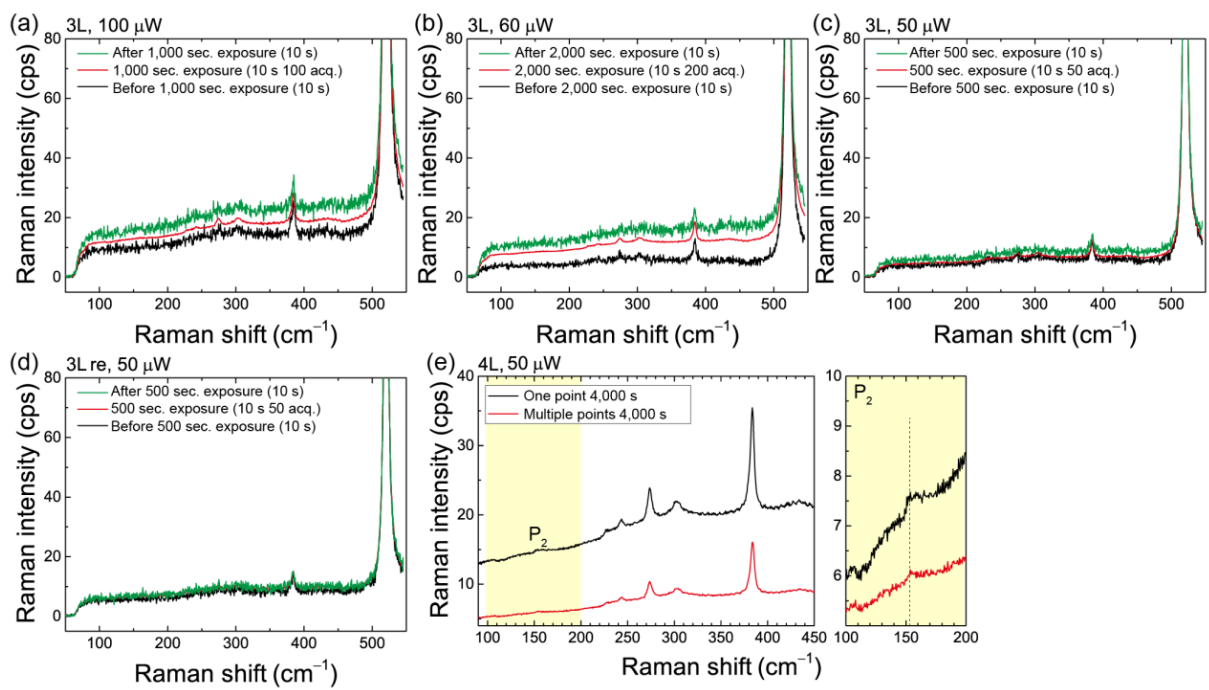

**Fig. S1.** Laser-induced degradation of few-layer MnPS$_3$. The excitation source is a 488-nm laser. For each case, a Raman spectrum was measured from a fresh spot with a short integration time of 10 s before a longer acquisition of 500 to 2000 s. Another Raman spectrum with a short integration time (10 s) was measured to assess the laser-induced changes. Typically, additional luminescent signal causes an increase of the baseline signal when the sample is damaged. Comparison of 3L MnPS3 samples exposed to a laser beam for (a) 1,000 sec with the power of 100 μW, (b) 2,000 sec with 60 μW, and (c) 500 sec with 50 μW. (d) Repeat of the measurements of (c) at the same spot. (e) Comparison of a spectrum taken at a single position for with an integration time of 4000 s and one obtained by adding spectra taken from 40 different points with an integration time of 100 s each.

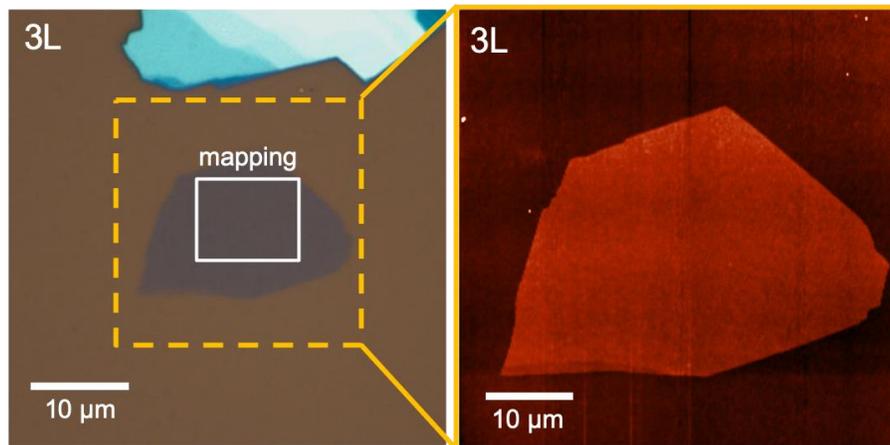

**Fig. S2.** Optical and AFM images of 3L MnPS$_3$. The dashed-yellow box indicates the region where the AFM image was taken. The white-solid box indicates the region where Raman spectra in Fig. S3 were taken.

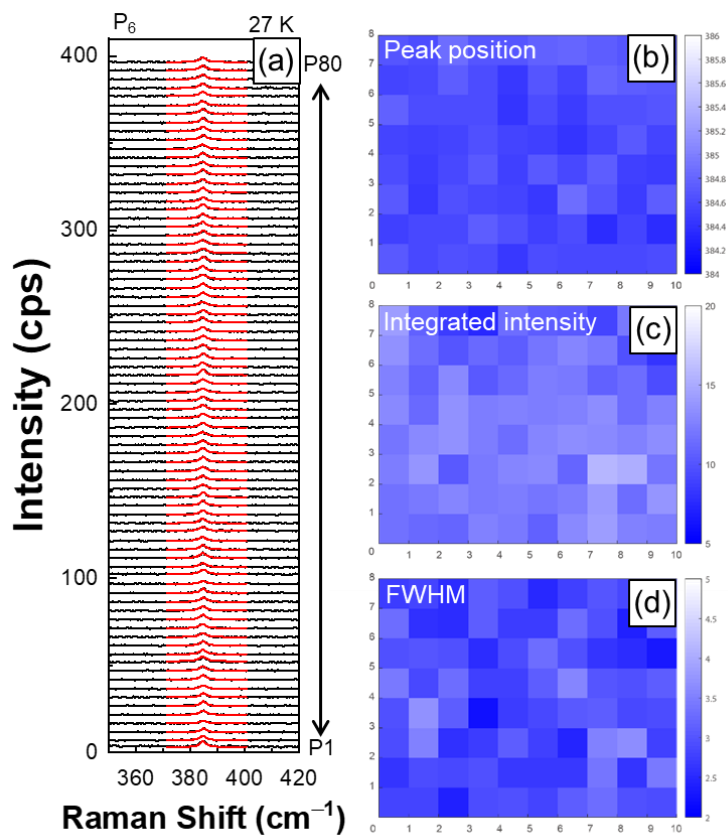

**Fig. S3**. (a) Raman spectra 3L MnPS$_3$ at 27K taken from the area indicated in Fig. S2(a). Raman maps of P$_6$ from the spectra in (a): (b) peak position, (c) integrated intensity, and (d) full width at half maximum (FWHM).

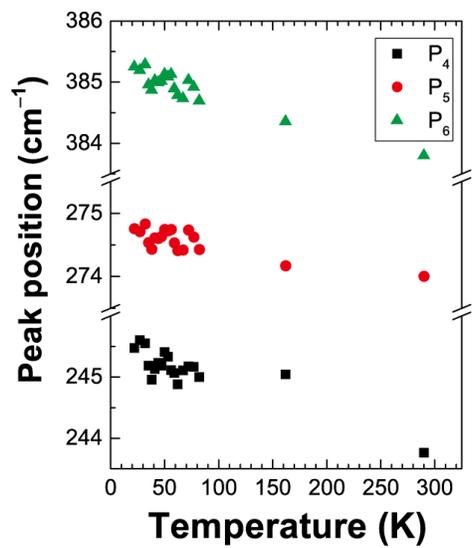

**Fig. S4**. Temperature-dependent peak positions of $P_4$, $P_5$, and $P_6$ of 3L $MnPS_3$.